\documentclass[twocolumn,amsmath,amssymb,aps,pre]{revtex4-1}

\usepackage{graphicx}
\usepackage{dcolumn}
\usepackage{bm}
\usepackage{xcolor}
\newcommand{\be}{\begin{equation}}
\newcommand{\ee}{\end{equation}}
\newcommand{\bea}{\begin{eqnarray}}
\newcommand{\eea}{\end{eqnarray}}

\newcommand{\sinf}{{s}}
\newcommand{\cosf}{{c}}

\newcommand{\tcb}[1]{\textcolor{black}{#1}}
\newcommand{\cor}[1]{\textcolor{black}{#1}}
\newcommand{\er}[1]{\textcolor{black}{#1}}
\newcommand{\el}[1]{\textcolor{black}{#1}}
\usepackage{mathtools}
\begin{document}

\title{Local master equations may fail to describe dissipative critical behavior}

\author{Michael Konopik}
\author{Eric Lutz}
\affiliation{Institute for Theoretical Physics I, University of Stuttgart, D-70550 Stuttgart, Germany}
\begin{abstract}
Local quantum master equations provide a simple description of interacting subsystems coupled to different reservoirs. They have been widely used to study nonequilibrium critical phenomena in open quantum systems. We here investigate the validity of such a local approach by analyzing a paradigmatic system made of two harmonic oscillators each in contact  with a heat bath. We evaluate the steady-state mean occupation number for varying  temperature differences and find that local master equations generally fail  to reproduce the results of an exact quantum-Langevin-equation description. We relate this property to  the inability of the local scheme to properly characterize intersystem correlations, which we quantify with the help of the quantum mutual information. 
 \end{abstract}

\maketitle
\section{Introduction}
Quantum master equations have been instrumental in the study of  open quantum systems  since their introduction by  Wolfgang Pauli in 1928 \cite{pau28}. They offer  powerful, yet approximate, means to describe the time evolution of the reduced density operator of  quantum systems coupled to  external environments \cite{bre02,gar04,ali07,riv12}. They allow the analysis of the dynamics of both  diagonal density matrix elements  (populations), involved in thermalization processes, and of  nondiagonal density matrix elements  (coherences), associated with dephasing phenomena.  As a consequence, they have found widespread application in many different areas,  ranging from  quantum optics \cite{car93} and condensed matter physics \cite{wei08} to nonequilibrium statistical mechanics \cite{zwa01} and quantum information theory \cite{nie00}.

In the past decade, quantum master equations have become a popular tool to investigate nonequilibrium  phase transitions that occur between  (detailed-balance breaking) steady states \cite{die08,die10,die08,bre13,car13,mar14,lab16,sor18,pro08,kar09,car09,pro10,pro11,pro11a,pro11b,vog12,cui15,fos17,car19,pop20}. Special attention has been given to two broad classes of out-of-equilibrium phase transitions: (i) those induced by external driving fields in  systems interacting with a single bath (driven-dissipative processes) \cite{die08,die10,bre13,die08,car13,mar14,lab16,sor18} and (ii) those generated by the coupling of a system to several baths (boundary-driven processes) \cite{pro08,kar09,car09,pro10,pro11,pro11a,pro11b,vog12,cui15,fos17,car19,pop20}.
  Remarkably,  nontrivial exact analytical steady-state solutions of local quantum Lindblad master equations  have  been  obtained for various many-body  spin-chain models \cite{pro08,kar09,pro11,pro11a,pro11b,vog12,fos17,pop20}, thus offering new insight into  boundary-driven critical systems. 

However, the form of the quantum master equations employed in  these studies is often postulated. Their validity is thus not completely clear a priori. This is especially true for boundary-driven processes where the system of interest is coupled to several reservoirs. In this case, it has recently been  shown that local master equations, that are commonly used to examine nonequilibrium phase transitions \cite{pro08,kar09,car09,pro10,pro11,pro11a,pro11b,vog12,cui15,fos17,car19,pop20}, may violate the second law of thermodynamics \cite{lev14} and give rise to nonphysical results, such as incorrect steady-state distributions or nonzero currents for vanishing bath interactions  \cite{wal70,car73,man15,tru16,gon17,hof17,sto17,nas18,chi17,mit18,cat19}, even in the limit of small bath couplings. These inconsistencies are related to the fact that local quantum master equations, whose total dissipator is simply the sum of the single-bath dissipators, incorrectly neglect bath-bath  correlations, which are induced by  intersystem interactions, in contrast to global  quantum master equations \cite{lev14,wal70,car73,man15,tru16,gon17,hof17,sto17,nas18,chi17,mit18,cat19}. Interestingly, the local approach has been shown to provide a better description of quantum heat engines than the global approach in some parameter regimes  \cite{hof17}. Meanwhile, the validity of Lindblad quantum  master equations has, for example, been discussed   in the context of quantum transport \cite{wic07,pur16}, quantum relaxation \cite{riv10,boy17},  and entanglement generation \cite{lud10}. But these results cannot be straightforwardly extended to nonequilibrium phase transitions as the considered models do not exhibit critical behavior.
\begin{figure}[t]
\includegraphics[width=0.36\textwidth]{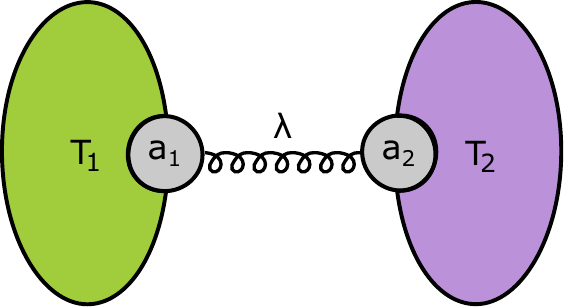}
\caption{Coupled-oscillator model. Two quantum harmonic oscillators interact with each other with interaction  strength $\lambda$. Each of them is weakly coupled with a heat bath with respective temperature $T_j$, $(j=1,2)$. A nonequilibrium steady state is established when the two temperatures are different and heat flows from one oscillator to the other.} 
\label{setup}
\end{figure}

In this paper, we examine the accuracy of a quantum-master-equation description of dissipative critical phenomena by analyzing an exemplary  system consisting of two interacting harmonic oscillators, each weakly coupled to a thermal reservoir. This system naturally appears in many areas, most notably in cavity optomechanics \cite{asp14}. \el{Many-body} superradiant phase-transition models, such as the Dicke model \cite{dic54} and the Tavis-Cummings model \cite{tav68}, can also be mapped onto such a system after a Holstein-Primakoff transformation \cite{bra05,kir19}. We concretely compare local and global quantum master equations, with and without rotating-wave approximation for the oscillator-oscillator interaction, to exact results provided by a quantum-Langevin-equation description \cite{pur16,riv10,boy17,lud10}. We explicitly evaluate the stationary mean occupation number of one  oscillator  for various nonequilibrium temperature differences. We find that the local master equation generally fails to reproduce the results of the quantum Langevin equation  especially  for large temperature differences, while the global approach exhibits better agreement. We show that this feature is directly related to the inability of the local description to correctly capture intersystem correlations, which we quantify with the help of the quantum mutual information \cite{nie00}.

\section{Coupled-oscillator model} 

We consider a system of two interacting  harmonic oscillators with Hamilton operator,
\begin{equation}
{H} =  \sum_{j=1, 2}\omega_j {a}_j^\dagger{ a}_j  +  \lambda ({a}_1 {a}_2^\dagger +{ a}_1^\dagger {a}_2)+ \kappa ({a}_1 {a}_2 +{ a}_1^\dagger {a}_2^\dagger),
\label{Hamiltonian}
\end{equation}
where ${a}_j^\dagger$ and ${ a}_j$  are the usual ladder operators and $\omega_j$ the respective frequencies. We will examine two different  \cor{types of intersystem} interactions:  (i) a   position-position interaction, $x_1 x_2$, corresponding to $\kappa=\lambda$, and  (ii) its rotating-wave version, obtained for  $\kappa=0$. 
Two important points should be stressed: First, the position-position coupling $x_1 x_2$ leads to critical behavior above a critical interaction strength \cite{ema03,lam04,sud12}, in contrast to the commonly treated Hookian interaction $(x_1-x_2)^2$  \cite{wic07,pur16,riv10,boy17,lud10}. In addition, while the rotation-wave approximation is usually associated with a weak-coupling condition, $\lambda/\omega_i \ll 1$, it has recently been shown that counter-rotating terms may be effectively suppressed in modulated systems, even  in the ultrastrong regime \cite{hua20,for19}. This opens the possibility to experimentally study critical behavior in strongly interacting rotating-wave models.

The isolated Hamilton operator (1) may be diagonalized exactly for both \cor{intersystem} interactions, yielding two uncoupled modes with respective energies \cite{ema03,lam04,lev14},
\begin{eqnarray}
\omega^\text{pp}_\pm &=& \left[(\omega_1 ^2 + \omega_1^2 \pm \sqrt{(\omega_1^2 - \omega_2^2)^2 + 16 \lambda^2 \omega_1 \omega_2})/2\right]^\frac{1}{2}\!, \\
\omega^\text{rw}_\pm &=& (\omega_1 + \omega_1 \pm \sqrt{(\omega_1 - \omega_2)^2 + 4 \lambda^2})/2.
\label{e2}
\end{eqnarray}
These energies display critical behavior at the respective critical couplings  $\lambda_\text{c}^\text{pp}=\sqrt{\omega_1 \omega_2}/2$ and $\lambda_\text{c}^\text{rw}=\sqrt{\omega_1 \omega_2}$. Above these points, the eigenfrequencies of the Hamilton \cor{operator (1)} become imaginary or negative. \cor{The energy spectrum is thus} no longer bounded from below. These \cor{critical} values are in agreement with those of the Dicke and Tavis-Cummings models \cite{hep73,hio73,car73a}.

We next attach each quantum harmonic oscillator to a heat bath with respective temperature $T_j$ (Fig.~\ref{setup}). As commonly done, we model these reservoirs by  an ensemble of harmonic oscillators \cite{bre02,gar04,ali07,riv12}. We further assume that the system-bath coupling is weak, so that the rotating-wave approximation is applicable to that coupling (\cor{see details in the Supplemental Material \cite{sup}}). We reemphasize that the interaction between the two  harmonic oscillators of the system (1) might be strong.

\er{Most studies of dissipative phase transitions  consider complex interacting many-body systems \cite{pro08,kar09,car09,pro10,pro11,pro11a,pro11b,vog12,cui15,fos17,car19,pop20}. The direct comparison between global and local master equation descriptions is thus extremely difficult in these systems. By contrast, the  coupled-oscillator model is complicated enough  to exhibit steady-state critical behavior and, at the same time, simple enough to  allow for (i) a detailed comparison between global and local approaches, and (ii)  the  evaluation of intersystem correlations. }

 \begin{figure}[t]
\includegraphics[width=0.49\textwidth]{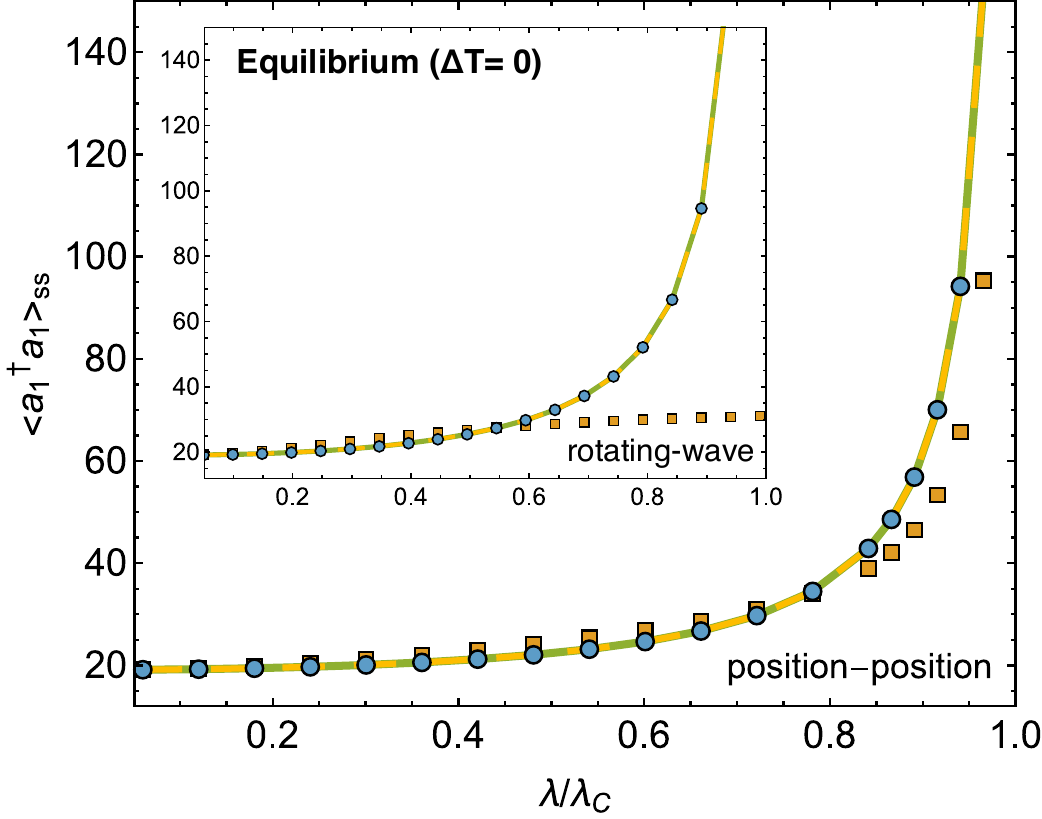}

\caption{Steady-state mean occupation number $\langle a_1^\dagger a_1\rangle_\text{ss}$ of the first  oscillator as a function of the dimensionless interoscillator interaction strength, $\lambda/\lambda_\text{c}$, for the  equilibrium (high-temperature) case $\Delta T=0$. For the position-position interaction [see Eq.~(1)], the results of the global quantum master equation (blue dots) perfectly agree with those of the quantum Langevin equation (green line) \cor{as well as those of the Gibbs state $\rho_\text{eq}=\exp(-\beta H)/Z$ (yellow line)}, while those of the local quantum master equation deviate more and more as the critical point is approached. For the rotating-wave interaction (inset), the global approach still perfectly matches the predictions of the quantum Langevin equation, while the local scheme does not display any critical behavior. Parameters are $\gamma_1=\gamma_2=1.5\cdot 10^{-4}$, $\omega_1=5$, $\omega_2=2$ and  $T_1=T_2=98$.}
\label{f1}
 \end{figure} 
 
\section{Quantum-master-equation description} In the usual Born-Markov limit, the density operator  $\rho$ of the joint quantum system obeys a Lindblad master equation of the form \cite{bre02,gar04,ali07,riv12} (we set $\hbar=1$ throughout),
\begin{equation}
\dot\rho= -i[H,\rho] + \sum_{k=1,2}\sum_{i= 1,2,3,4}  \sum_{j= 1,2,3,4}\mathcal{D}_{k}({A_i,A_j}),
\label{genME}
\end{equation}
where  the dissipators are given by $\mathcal{D}_{k}({A_i,A_j})=\Gamma_k({A_i,A_j}) (A_i \rho A_j - \{ A_j A_i, \rho\}/2)$. The  coefficients $\Gamma_k({A_i,A_j})$, as well as the operators $A_i$,  depend on the local or global type of the quantum master equation \cite{lev14}.

In the local approach, each  oscillator interacts with its heat bath \el{(labelled by $k=1$ or $2$)} as if it were not coupled to the other oscillator. As a result, the quantum master equation may be derived as usual in the local eigenbasis of one oscillator \cite{bre02,gar04,ali07,riv12}. The operators $A_i$ are here the standard ladder operators,  $ (a_1,a_1^\dagger,a_2,a_2^\dagger)$, and the  dissipators are given by  $\Gamma_{k}(a_i, a_j^\dagger)=\delta_{kj} \delta_{ij}\gamma_k [N(\omega_j, \beta_k)+1]$ and $\Gamma_{k}({a_i^\dagger, a_j}) =\delta_{kj} \delta_{ij} \gamma_k N (\omega_j, \beta_k)$, where $\gamma_k$ is the damping coefficient \el{of bath $k$} and  $N(\omega_j, \beta_k)= 1/[\exp(\beta_k \omega_j)-1]$  denotes the thermal occupation number \cite{bre02,gar04,ali07,riv12}. These formulas evidently hold for the two kinds of oscillator-oscillator interaction in Eq.~(1).

On the other hand, the global master equation is derived in the global eigenbasis of the combined two-oscillator system \cite{riv10,lev14}. The diagonalization of the joint Hamilton operator (1) accounts for the indirect subsystem-reservoir and reservoir-reservoir   correlations which are generated by their coupling to the system. Such correlations are ignored in the local approach. This is the reason why the local master equation may violate the second law of thermodynamics \cite{lev14}. The explicit (and lengthy) expressions for the dissipators are summarized for both the position-position and rotating-wave interactions in the Supplemental Material \cite{sup}. In this situation, they  depend  on operators $A_i$  that are given by   properly rotated ladder operators \cite{sup}. 

In the following, we will  solve the four different quantum master equations (global/local forms with/without rotating-wave interaction) by applying a characteristic function method in symplectic space \cite{sup} and evaluate the steady-state mean occupation number   $\langle a_j^\dagger a_j\rangle_\text{ss} = \text{tr}(\rho_\text{ss} a_j^\dagger a_j)$, \cor{where $\rho_\text{ss}$ is the stationary density operator}.

 \begin{figure}[t]
\includegraphics[width=0.48\textwidth]{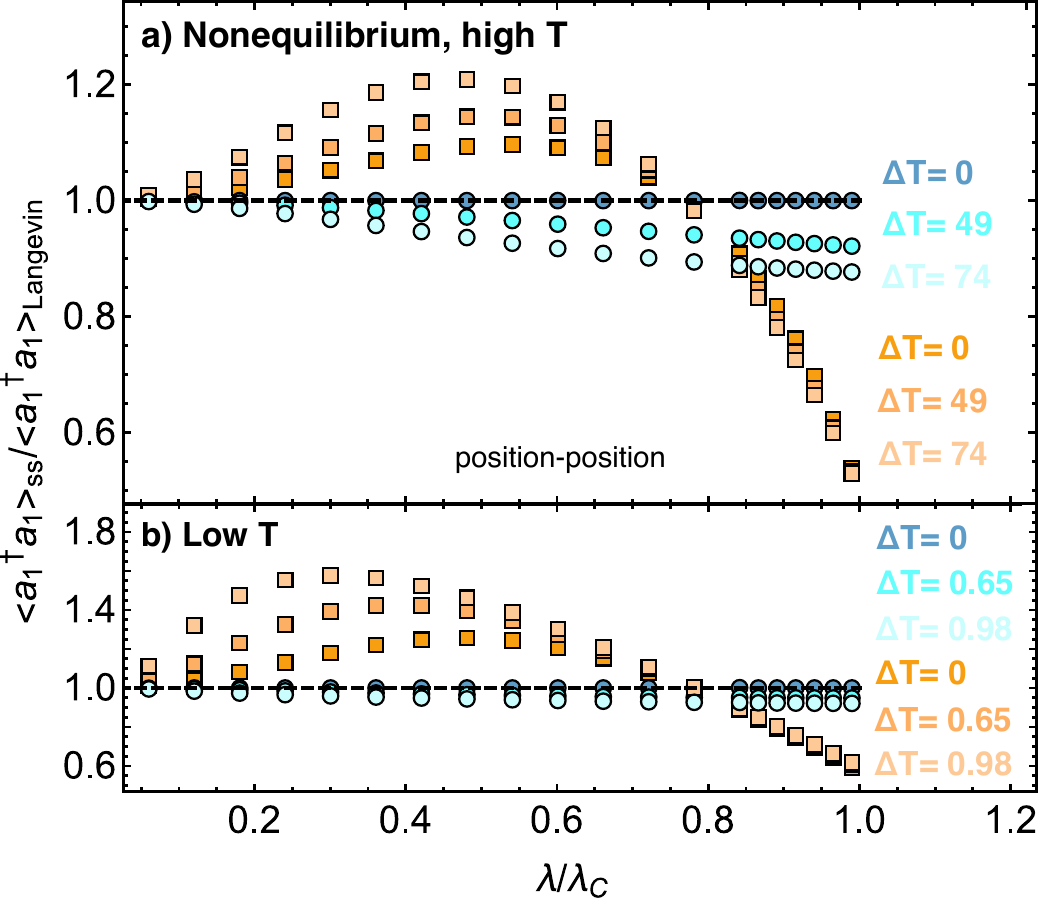}

\caption{Ratio  of the steady-state mean occupation numbers $\langle a_1^\dagger a_1\rangle_\text{ss}/ \langle a_1^\dagger a_1\rangle_\text{Langevin}$ of the quantum master equation and the quantum Langevin equation as a function of $\lambda/\lambda_\text{c}$ for various nonequilibrium temperature differences $\Delta T$ and position-position interaction [see Eq.~(1)]. a) In the high-temperature regime ($\beta_i \omega_i\ll 1$), \cor{the  local (global)  quantum master equation [orange squares (blue dots)] strongly (slightly)  departs from the  predictions of the quantum Langevin  equation  for increasing temperature differences, missing the critical behavior for all $\Delta T$. }b) \cor{An analogous behavior is observed} for  low temperatures ($\beta_i \omega_i\gg 1$). Same parameters as in Fig.~\ref{f1}.}
\label{f2}
 \end{figure} 
 
 \begin{figure}[t]
\includegraphics[width=0.47\textwidth]{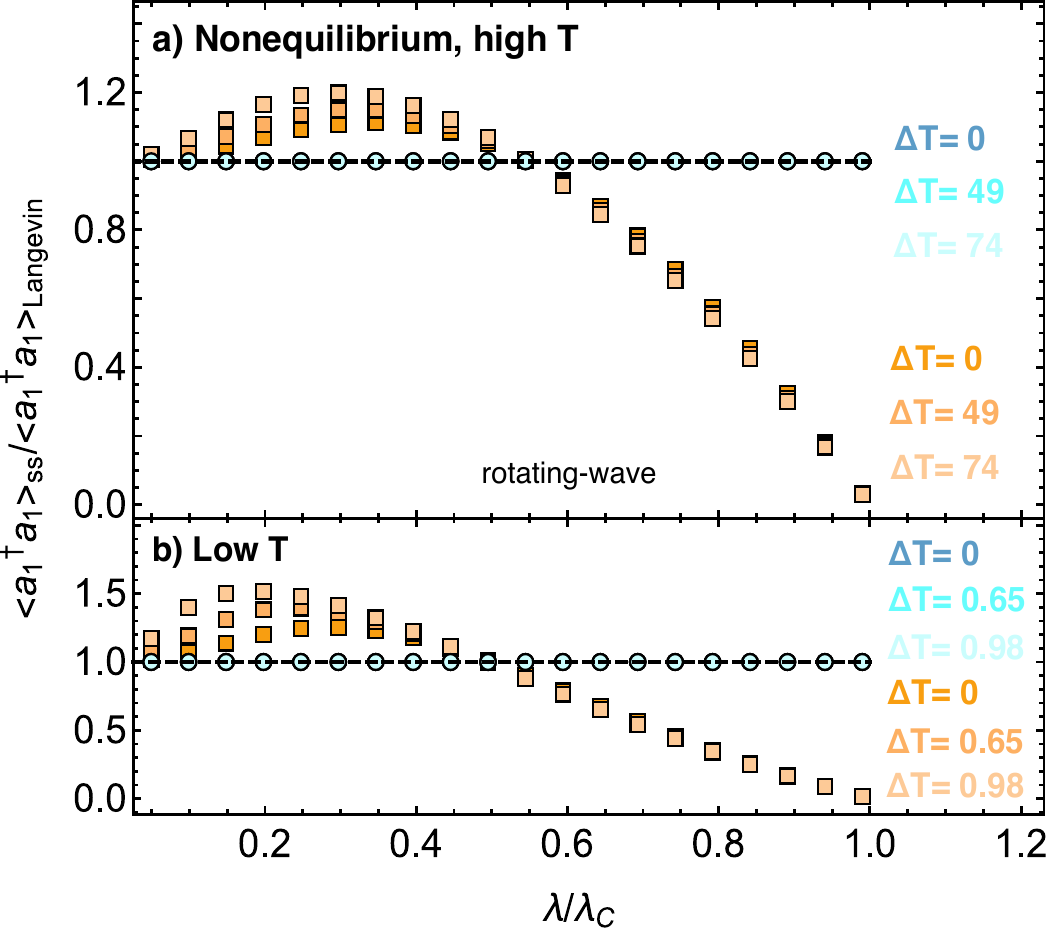}

\caption{Ratio  of the steady-state mean occupation numbers $\langle a_1^\dagger a_1\rangle_\text{ss}/ \langle a_1^\dagger a_1\rangle_\text{Langevin}$ of the quantum master equation and the quantum Langevin equation as a function of $\lambda/\lambda_\text{c}$ for various  temperature differences $\Delta T$ and rotating-wave interaction [see Eq.~(1)]. Both for a)  high temperatures  ($\beta_i \omega_i\ll 1$) and b) low temperatures ($\beta_i \omega_i\gg 1$),  the global  (blue dots) quantum master equation exactly agrees with the  quantum Langevin  equation, while the local quantum master equation (orange squares) shows large deviations for  increasing temperature differences.  Same parameters as in Fig.~\ref{f1}, \cor{with $T_2 = 1.96$ and $T_1 = T_2 - \Delta T$ in the low-temperature limit}.}
\label{f3}
 \end{figure}

\section{Quantum-Langevin-equation description} In order to assess its validity, \cor{both for equilibrium and nonequilibrium conditions}, we shall compare the steady-state properties of the approximate quantum-master-equation treatment to those of the exact quantum-Langevin-equation approach \cite{gar04}. To this end, we will extend the results  obtained for Hookian coupling \cite{pur16,riv10,boy17,lud10} to  the position-position and rotating-wave interactions of Eq.~(1).
The quantum Langevin equations read \cite{gar04},
\begin{equation}
\label{5}
\dot{ a}_j = -i[a_j,H]-\gamma_j a_j + \sqrt{2 \gamma_j} a_{j,\text{in}},
\end{equation}
where the noisy input operators $a_{j,\text{in}}$\tcb{, stemming from the interaction with the respective baths,} are characterized  by the correlation function in Fourier space,
\begin{eqnarray}
\langle a_{j,\text{in}}(\nu_j) a_{j,\text{in}}^\dagger(\nu_j') \rangle &=& 2 \gamma_j[N(\nu_j,\beta_j)+1]\delta(\nu_j-\nu_j'), \\
\langle a_{j,\text{in}}^\dagger(\nu_j')  a_{j,\text{in}}(\nu_j)  \rangle &=& 2 \gamma_j N(\nu_j,\beta_j) \delta(\nu_j-\nu_j'),
\label{inmean}
\end{eqnarray}
The coupled quantum Langevin equations \eqref{5} can be solved by matrix inversion in Fourier space \cite{sup}. In particular, 
the mean occupation number is here equal to, 
\begin{equation}
\langle a_j^\dagger a_j\rangle_\text{Langevin}=\int_{-\infty}^\infty  \int_{-\infty}^\infty \langle a_j^\dagger (\nu_j) a_j(\nu_j')\rangle e^{i (\nu_j - \nu_j') t} d\nu_j  d\nu_j'.
\label{Lanlsg}
\end{equation}
 \cor{Equation \eqref{Lanlsg} is independent of time in the steady-state regime and we will set $t=0$ in the following.} \el{Steady-state mean occupation numbers may be evaluated exactly (without any approximations) in the quantum-Langevin-equation formalism  in contrast to the quantum-master-equation approach \cite{pur16}.}
   
\section{Results}  Figure 2 presents the steady-state mean occupation number $\langle a_1^\dagger a_1\rangle_\text{ss}$ of the first  oscillator as a function of the reduced interaction strength $\lambda/\lambda_\text{c}$ in the equilibrium (high-temperature) case $\Delta T = T_2-T_1=0$. We observe perfect agreement between the global quantum master equation (blue dots), the quantum Langevin equation (green line) \cor{and the equilibrium (Gibbs) state $\rho_\text{eq} = \exp(-\beta H)/Z$ (yellow line) \cite{sup}} for all values of $\lambda/\lambda_\text{c}$, for both the position-position and the rotating-wave (inset) interactions. By contrast, the local quantum master equation (orange squares) deviates from these results as the critical point is approached; \cor{noticeably,} it does not exhibit any critical behavior at all for the \cor{intersystem} rotating-wave interaction (inset).

In order to gain deeper insight on the nonequilibrium properties of the different quantum master equations, we next examine the ratio of their steady-state mean occupation numbers and the corresponding quantum-Langevin-equation expressions, $\langle a_1^\dagger a_1\rangle_\text{ss}/\langle a_1^\dagger a_1\rangle_\text{Langevin}$, for increasing temperature differences $\Delta T$.  In the high-temperature  regime ($\beta_i \omega_i\ll 1$), Fig.~3a shows that, for the position-position interaction, the local approach gets worse as the system moves further away from equilibrium and that even the global approach slightly departs from the predictions of the quantum Langevin equation for large $\Delta T$. \cor{A similar behavior is seen in Fig.~3b} when the two  temperatures are low ($\beta_i \omega_i\gg 1$). Analogous results are displayed for the rotating-wave interaction in Figs.~4ab: remarkably, the global quantum master equation here always perfectly matches the quantum Langevin equation, for all $\lambda$ and all $\Delta T$, while the local  quantum master equation always fails to describe critical behavior. We also mention that the discrepancy between the various descriptions in general depends on the sign of the nonequilibrium temperature difference $\Delta T$ \cite{sup}.

 \begin{figure}[t]
\includegraphics[width=0.49\textwidth]{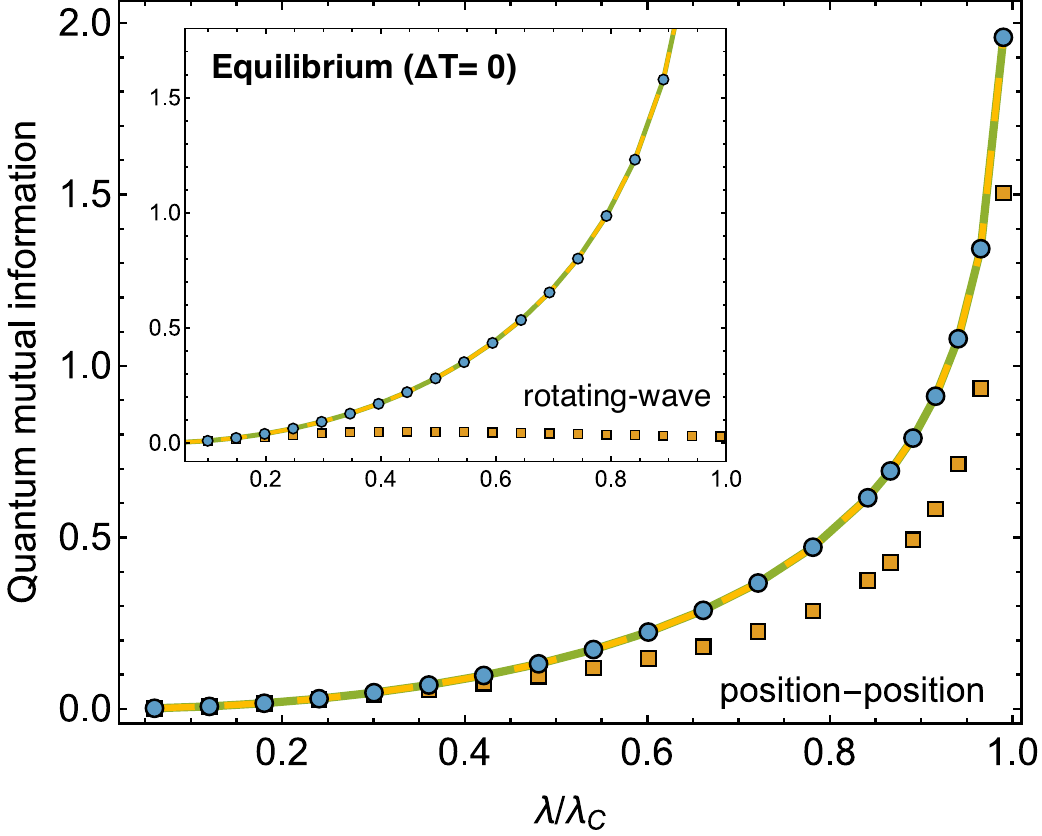}

\caption{Steady-state quantum mutual information  $I(\rho_\text{ss})$  of the two-oscillator system as a function of the dimensionless interoscillator interaction strength, $\lambda/\lambda_\text{c}$, for the  equilibrium (high-temperature) case $\Delta T=0$. The mutual information displays an analogous dependence of the interaction strength as the steady-state mean occupation number $\langle a_1^\dagger a_1\rangle_\text{ss}$ shown in Fig.~1. Same parameters as in Fig.~\ref{f1}.}
\label{f4}
 \end{figure} 
 
The success/failure of the quantum-master-equation description of dissipative critical phenomena may be understood both physically and mathematically. To first address the physical aspect, we consider the quantum mutual information between the two harmonic oscillators, $I(\rho)= S(\rho_1) + S(\rho_2) - S(\rho)$, where $S(\rho_i)=- \text{tr}\{\rho_i\ln \rho_i \}$ is the von Neumann entropy and $\rho_i = \text{tr}_i \rho$ are the reduced density operators of the respective harmonic oscillators \cite{nie00}. The quantum mutual information is a measure of the total (classical and quantum) correlations between two subsystems and has been used broadly to characterize critical transitions \cite{anf05,sin11,alc13,tom17,wal19}. Figure 5 shows that the stationary quantum mutual information $I(\rho_\text{ss})$ displays a very similar dependence on the  interaction strength $\lambda$ as the average  occupation number $\langle a_1^\dagger a_1\rangle_\text{ss}$ represented in Fig.~1, both for the position-position and rotating-wave interoscillator interactions. The shortcomings of the quantum-master-equation approach, especially in its local version, may thus be traced to its inability to correctly capture intersystem correlations close to the critical point. This feature can be confirmed mathematically by looking at the way  the respective Lindblad quantum master equations are obtained \cite{lev14}: the dissipators in the local master equation are indeed derived in the local eigenbasis of each separate harmonic oscillator, while those of the global master equation follow from a diagonalization of the interacting two-oscillator system (the unitary evolution given by the von Neumann term in Eq.~(4) describes coupled dynamics in both cases). The global scheme  thus better accounts for intersystem correlations than the local one, \el{and should therefore be preferred. Such intersystem correlations are indeed crucial for an accurate description of many-body critical systems, and should not be incorrectly omitted} Yet, despite these deficiencies, local quantum master equations have been a tool of choice in numerous  studies on dissipative critical behavior 
\cite{pro08,kar09,car09,pro10,pro11,pro11a,pro11b,vog12,cui15,fos17,car19,pop20}.

\section{Conclusions} We have examined the ability of global and local quantum master equations to accurately describe dissipative critical phenomena using an illustrative system of two interacting, damped harmonic oscillators, with and without rotating-wave interaction. \er{This model provides a transparent, yet generic, example  to perform such a study.} We have found that while the global master equation  reproduces  the results of the quantum Langevin equation reasonably well, the local version usually fails to do so, especially in the far-from-equilibrium regime; it  generally fails in the case of  the rotating-wave interaction. We have related these properties to the inability of the local approach to correctly apprehend oscillator-oscillator correlations that we have quantified with the help of the quantum mutual information. \er{The latter quantity could be easily determined in the present two-oscillator model, in contrast to more complex interacting many-body systems.} Our findings show that approximate local quantum master equations in general, and their exact analytical solutions in particular, should be used with caution when studying dissipative critical behavior, \el{and that the more complicated global approach should be favored instead}.

\section*{Acknowledgments}
We acknowledge   financial support from the German Science Foundation (DFG) (Contract No FOR 2724).

\section*{Appendix A: Quantum master equations}
In the following, we provide  details about  the dissipators  of the four quantum master equations  that we consider in our study (global/local forms with/without rotating-wave interaction) as well as their solutions.

The standard dissipators of the local quantum master equation are given below Eq.~\eqref{genME} in the main text. They are derived in the local eigenbasis of each oscillator \cite{bre02,gar04,ali07,riv12} and thus hold for both the position-position and rotating-wave intersystem interactions. By contrast, the global master equations are  derived in the global eigenbasis of the combined two-oscillator system obtained by diagonalizing the quadratic Hamilton operator $H$. The respective dissipators are then computed  by expanding the system-bath interaction in this basis.  We concretely consider the total  system-bath Hamilton operator,
\begin{equation}
H_\text{tot} = H + H_{SB1}+ H_{SB2}+ H_{B1} + H_{B2},
\end{equation}
with harmonic thermal baths $H_{Bi} = \sum_j \omega_{ij} b_{ij}^\dagger b_{ij}$ and local system-bath couplings $H_{SBi} = \sum_j\kappa_{ij} (a_i b_{ij}^\dagger +  \text{h.c.})$ with coupling constants $\kappa_{ij}$ ($i=1, 2$) \cite{bre02,gar04,ali07,riv12}.

In the case of the rotating-wave  interaction, the diagonalization of $H$  leads to  $H_\text{rw} = \omega_+^\text{rw} d^\dagger_+ d_+ +\omega_-^\text{rw} d^\dagger_- d_- $ with the eigenfrequencies $\omega_\pm^\text{rw} $ given in  Eq.~(3) of the main text and the rotated operators $d_- = a_2 \cos\theta - a_1 \sin\theta$ and $d_+ = a_1 \cos\theta + a_2 \sin\theta$, where  the angle $\theta$ satisfies $\cos^2\theta={(\omega^\text{rw}_+-\omega^\text{rw}_-)}/{(\omega^\text{rw}_+-\omega^\text{rw}_-)}$ \cite{lev14}. 
The global Lindblad dissipators then follow as \cite{lev14},
\begin{eqnarray}
\Gamma(a_1, a_1^\dagger)&=& \gamma_1^+ \cosf^4+ \gamma_1^- \sinf^4 + (\gamma_2^+ +\gamma_2^-) \cosf^2 \sinf^2 \\
\Gamma(a_1^\dagger, a_1)&=& \gamma_1^{+}{}' \cosf^4 + \gamma_1^{-}{}' \sinf^4 + (\gamma_2^{+}{}'  +\gamma_2^{-}{}' ) \cosf^2 \sinf^2  \\
\Gamma(a_2, a_2^\dagger)&=& \gamma_2^- \cosf^4+ \gamma_2^+ \sinf^4 + (\gamma_1^+ +\gamma_1^-) \cosf^2 \sinf^2 \\
\Gamma(a_2^\dagger, a_2)&=& \gamma_2^+{}' \cosf^4 + \gamma_2^-{}' \sinf^4  + (\gamma_1^+{}'  +\gamma_1^-{}' ) \cosf^2 \sinf^2  \\
\Gamma(a_1, a_2^\dagger)&=& \gamma_1^+ \cosf^3\sinf- \gamma_1^- \sinf^3\cosf + \gamma_2^+ \cosf \sinf^3- \gamma_2^-\cosf^3 \sinf \\
\Gamma( a_1^\dagger, a_2)&=&  \gamma_1^+{}' \cosf^3\sinf  - \gamma_1^-{}' \sinf^3\cosf  + \gamma_2^+{}' \cosf \sinf^3 - \gamma_2^-{}'\cosf^3 \sinf,
\label{Tavisme}
\end{eqnarray}
with $c=\cos\theta$, $s=\sin \theta$, $\gamma_{i}^\pm = \gamma$, $\gamma_i ^{\pm}{}'= \gamma_{i}^\pm e^{-\beta_i \omega_\pm}$
and $\Gamma( a_1^\dagger, a_2)= \Gamma
(a_2^\dagger, a_1),~ \Gamma(a_1, a_2^\dagger)=\Gamma(a_2, a_1^\dagger)$.

In the case of the position-position interaction, the diagonalization of $H$ is more involved as it couples all four ladder operators with each other, $(a_1,a_2,a_1^\dagger, a_2^\dagger)= S (c_1,c_2,c_1^\dagger, c_2^\dagger)$ \cite{ema03}. The $4\times 4$ diagonalization matrix $S$ is partitioned  into four blocks with the $2\times 2$ matrix $A$ on the diagonal blocks and   $2\times 2$ matrix $B$ matrix on the off-diagonal blocks:
\begin{equation}
\begin{aligned}
A&=\left(\begin{matrix}
\frac{(\omega^\text{pp}_+ + \omega_1)\cos\theta}{2\sqrt{\omega^\text{pp}_+ \omega_1}}&\frac{-(\omega^\text{pp}_- + \omega_1)\sin\theta}{2\sqrt{\omega^\text{pp}_- \omega_1}}\\
\frac{(\omega^\text{pp}_+ + \omega_2)\sin \theta}{2\sqrt{\omega^\text{pp}_+ \omega_2}}&\frac{-(\omega^\text{pp}_- + \omega_2)\sin\theta}{2\sqrt{\omega^\text{pp}_- \omega_2}}\\
\end{matrix}\right)\\
B&=\left(\begin{matrix}
\frac{(-\omega^\text{pp}_+ + \omega_1)\cos\theta}{2\sqrt{\omega^\text{pp}_+ \omega_1}}&\frac{(\omega^\text{pp}_- - \omega_1)\sin\theta}{2\sqrt{\omega^\text{pp}_- \omega_1}}\\
\frac{(-\omega^\text{pp}_+ + \omega_2)\sin\theta}{2\sqrt{\omega^\text{pp}_+ \omega_2}}&\frac{(\omega^\text{pp}_- - \omega_2)\sin\theta}{2\sqrt{\omega^\text{pp}_- \omega_2}}\\
\end{matrix}\right),
\end{aligned}
\end{equation}
with the eigenfrequencies $\omega_\pm^\text{pp} $ given in  Eq.~(2) of the main text. The global Lindblad dissipators are then $\Gamma(A_i, A_j)= \Gamma_{1,13}^{ij}+\Gamma_{1,31}^{ij} + \Gamma_{2,24}^{ij}+\Gamma_{2,42}^{ij}$, with
\begin{equation}
\begin{aligned}
\Gamma^{i j}_{1,k l}=& \gamma_1N(\omega_+,\beta_1) S_1^k S_3^l W_i^1 W_j^3\\
&+ \gamma_1N(\omega_-,\beta_1) S_2^k S_4^l W_i^2 W_j^4\\
&+ \gamma_1[N(\omega_+,\beta_1)+1] S_1^k S_3^l W_i^3 W_j^1\\
&+ \gamma_1[N(\omega_-,\beta_1)+1] S_2^k S_4^l W_i^4 W_j^2
\end{aligned}
\label{dickeme}
\end{equation}
for the quantum oscillator 1 coupled to bath 1 at inverse temperature $\beta_1$ with $W= S^{-1}$. Here the indexes $i,j$ run over 1-4, corresponding to the elements of $(a_1,a_2,a_1^\dagger, a_2^\dagger)$ and $k,l$ run over all combinations of 1,3. The indexes $k,l$ correspond to the initially chosen local coupling terms in the derivation of the master equation before the diagonalization is applied, which can be ordered either as $a_i a_i^\dagger$ or $a_i^\dagger a_i$. Thus, there are 32 different terms corresponding to the 16 unique operator orderings $A_i A_j$ in the dissipators. Expressions for second bath at inverse temperature $\beta_2$ are analogous with $k,l$ now combinations of 2,4.

We explicitly solve the linear local and global quantum master equations by computing the first and second moments of $\rho$ in  symplectic space \cite{bre02}.
The symmetric characteristic function is defined by $\chi (\alpha_1,\alpha_2)= \langle {D}_1(\alpha_1)\otimes {D}_2(\alpha_2) \rangle$, where ${D}_i(\alpha_i) = \exp( \alpha_i {a}_i^\dagger -  \alpha_i^* {a}_i )$ is  the displacement operator. The (symmetric) moments are then obtained by differentiation \cite{Cam16},
 \begin{equation}
\langle{a}_i^{\dagger k} {a}_j^l\rangle_s = \frac{d^k}{d\alpha_i^k} \frac{d^l}{(-\alpha_j^*)^l} \chi(\alpha_1,\alpha_2)|_{\alpha_1=\alpha_2 =0},
\label{symplectrafo}
\end{equation} 
{where $\langle\cdot\rangle_s$ is the expectation value of the symmetrized version of the operators $a_i^{\dagger k} a_j^l$.}
The evolution  of the characteristic function is derived from the  master equation
\begin{equation}
\frac{d}{dt}\chi(\alpha_1,\alpha_2) = \text{Tr}\{  {D}_1(\alpha_1)\otimes {D}_2(\alpha_2) \dot{\rho} \},
\end{equation}
together with the identities,
\begin{eqnarray}
{D}_i {a}_i^\dagger &=& \left(- \frac{\alpha_i^*}{2} + \frac{d}{d\alpha_i} \right){D}_i, \,{D}_i {a}_i = \left(- \frac{\alpha_i}{2}- \frac{d}{d\alpha_i^*} \right){D}_i, \nonumber \\
{a}_i^\dagger {D}_i  &=& \left( \frac{\alpha_i^*}{2} + \frac{d}{d\alpha_i} \right){D}_i, \,{a}_i {D}_i = \left( \frac{\alpha_i}{2} - \frac{d}{d\alpha_i^*} \right){D}_i.
\end{eqnarray}
with   $\alpha_i = x_i + i p_i$ and $d/d\alpha_i = (d/dx_i - i d/dp_i)/2$  using the Gaussian ansatz $\chi(x_1,p_1,x_2,p_2)  = \exp(i \vec P \vec{\bar{y}} -  \vec P^T \bar \sigma \vec P/2)$ with $\vec P = (x_1,p_1,x_2,p_2)^T$ and $\vec{\bar{y}} = (\bar{y}_1,\bar{z}_1,\bar{y}_2,\bar{z}_2)^T$.  Since the Hamiltonian is purely of quadratic order, the steady state values for the first moments always vanish $\bar{y}_i=0=\bar{z}_i$ and  the system is completely  described by the second moments. Writing these second moments in vector form $\vec{\sigma}=(\bar{\sigma}_{x1x1},\bar{\sigma}_{x1p1},\bar{\sigma}_{x1x2},\bar{\sigma}_{x1p2},\bar{\sigma}_{p1p1},\bar{\sigma}_{p1x2},$ $\bar{\sigma}_{p1p2},\bar{\sigma}_{x2x2},\bar{\sigma}_{x2p2},\bar{\sigma}_{p2p2})$, one may write the steady-state set of equations as $\vec{G}=\underline{\Lambda}\vec{\sigma}$.  

The $10\times 10$ matrix $\underline{\Lambda}$ can be written down row-wise using $2\Gamma(i,j,k,l,m,n,o,p)=(-1)^i {{\Gamma(a_1,a_2})+(-1)^j {\Gamma(a_1,a_2^{\dagger})}+(-1)^k{\Gamma(a_1^{\dagger},a_2}})+(-1)^l{\Gamma(a_1^{\dagger},a_2^{\dagger})}+ (-1)^m {\Gamma(a_2,a_1})+(-1)^n{\Gamma(a_2,a_1^{\dagger})}+(-1)^o{\Gamma(a_2^{\dagger},a_1})+(-1)^p {\Gamma(a_2^{\dagger},a_1^{\dagger})}$. We have
\begin{widetext}
\begin{equation}
\begin{aligned}
\vec{G}=&(~\Gamma (a_1,a_1)+\Gamma (a_1,a_1^{\dagger})+\Gamma (a_1^{\dagger},a_1)+\Gamma( a_1^{\dagger},a_1^{\dagger}),2 i \Gamma( a_1,a_1)-2 i \Gamma(a_1^{\dagger},a_1^{\dagger}),\Gamma(a_1,a_2)+\Gamma (a_1,a_2^{\dagger})+\Gamma(a_1^{\dagger},a_2)  \\
&+\Gamma( a_1^{\dagger},a_2^{\dagger})+\Gamma( a_2,a_1)+\Gamma( a_2,a_1^{\dagger})+\Gamma( a_2^{\dagger},a_1)+\Gamma( a_2^{\dagger},a_1^{\dagger}),
i (\Gamma (a_1,a_2)-\Gamma( a_1,a_2^{\dagger})+\Gamma (a_1^{\dagger},a_2)-\Gamma( a_1^{\dagger},a_2^{\dagger}) \\
&+\Gamma( a_2,a_1)+\Gamma (a_2,a_1^{\dagger})-\Gamma(a_2^{\dagger},a_1)-\Gamma( a_2^{\dagger}),a_1^{\dagger}),-\Gamma (a_1,a_1)+\Gamma( a_1,a_1^{\dagger})+\Gamma( a_1^{\dagger},a_1)-\Gamma( a_1^{\dagger},a_1^{\dagger}), \\
&i (\Gamma (a_1,a_2)+\Gamma( a_1,a_2^{\dagger})-\Gamma( a_1^{\dagger},a_2)-\Gamma( a_1^{\dagger},a_2^{\dagger})+\Gamma( a_2,a_1)-\Gamma( a_2,a_1^{\dagger})+\Gamma(a_2^{\dagger},a_1)-\Gamma( a_2^{\dagger},a_1^{\dagger})),-\Gamma( a_1,a_2) \\
&+\Gamma( a_1,a_2^{\dagger})+\Gamma(a_1^{\dagger},a_2)-\Gamma( a_1^{\dagger},a_2^{\dagger})-\Gamma (a_2,a_1)+\Gamma( a_2,a_1^{\dagger})+\Gamma(a_2^{\dagger},a_1)-\Gamma( a_2^{\dagger},a_1^{\dagger}),\Gamma(a_2,a_2)+\Gamma( a_2,a_2^{\dagger}) \\
&+\Gamma(a_2^{\dagger},a_2)+\Gamma( a_2^{\dagger},a_2^{\dagger}),2 i (\Gamma( a_2,a_2)-\Gamma( a_2^{\dagger},a_2^{\dagger})),-\Gamma(a_2,a_2)+\Gamma( a_2,a_2^{\dagger})+\Gamma( a_2^{\dagger},a_2)-\Gamma( a_2^{\dagger},a_2^{\dagger})~).\\
\underline{\Lambda}_1 =&
 ({\Gamma(a_1,a_1^{\dagger}})-{\Gamma(a_1^{\dagger},a_1)} ,  -\omega_1, \Gamma(1,0,1,0,0,0,1,1) , -i\Gamma(1,1,1,1,0,0,0,0)-(\kappa+\lambda),0,0,0,0,0,0)\\
 \underline{\Lambda}_2 =&
( \omega_1 ,  2 {\Gamma( a_1,a_1^{\dagger}})-2 {\Gamma (a_1^{\dagger},a_1}) , i\Gamma(1,0,0,1,0,1,1,0)-(\kappa - \lambda) , \Gamma(0,0,1,1,1,0,1,0) ,- \omega_1,\\
 & \Gamma(1,0,1,0,0,0,1,1) , i \Gamma(1,1,1,1,0,0,0,0)-(\kappa + \lambda),0,0,0)\\
  \underline{\Lambda}_3 =& (\Gamma(0,0,1,1,1,0,1,0),  i\Gamma(0,0,0,0,1,1,1,1) -(\kappa+\lambda), {\Gamma (a_1,a_1^{\dagger})}-{\Gamma (a_1^{\dagger},a_1)}+{\Gamma (a_2,a_2^{\dagger})}-{\Gamma (a_2^{\dagger},a_2)} ,\\
  & -\omega_2,0,-\omega_1,0, \Gamma(1,0,1,0,0,0,1,1),i\Gamma(1,1,1,1,0,0,0,0) -(\kappa+\lambda), 0)\\
  \underline{\Lambda}_4 =&( -(\kappa-\lambda) + i \Gamma(0,1,1,0,1,0,0,1), \Gamma(1,0,1,0,0,0,1,1), \omega_2,    {\Gamma (a_1,a_1^{\dagger})}-{\Gamma (a_1^{\dagger},a_1)}+{\Gamma (a_2,a_2^{\dagger})}-{\Gamma (a_2^\dagger,a_2)},\\
  & 0,0,-\omega_1,0,\Gamma(1,0,1,0,0,0,1,1), -(\kappa + \lambda) + i \Gamma(1,1,1,1,0,0,0)  )\\
  \underline{\Lambda}_5 =& (0,\omega_1,0,0,   {\Gamma (a_1,a_1^{\dagger})}-{\Gamma (a_1^{\dagger},a_1)}, - (\kappa-\lambda) + i \Gamma(1,0,0,1,0,1,1,0) , \Gamma(0,0,1,1,1,0,1,0),0,0,0)\\
 \underline{\Lambda}_6 =& ( 0,\Gamma(0,0,1,1,1,0,1,0), \omega_1,0, i\Gamma(0,0,0,0,1,1,1,1) -(\kappa + \lambda),    {\Gamma (a_1,a_1^{\dagger})}-{\Gamma (a_1^{\dagger},a_1)}+{\Gamma (a_2,a_2^{\dagger})}-{\Gamma (a_2^\dagger,a_2)}, \\
 &- \omega_2, i \Gamma(1,0,0,1,0,1,1,0) +(\lambda-\kappa) , \Gamma(0,0,1,1,1,0,1,0),0 )\\
 \underline{\Lambda}_7=&( 0,i \Gamma(0,1,1,0,1,0,0,1)+ (\lambda-\kappa), 0, \omega_1, \Gamma(1,0,1,0,0,0,1,1), \omega_2,{\Gamma (a_1,a_1^{\dagger})}-{\Gamma (a_1^{\dagger},a_1)}+{\Gamma (a_2,a_2^{\dagger})}-{\Gamma (a_2^\dagger,a_2)},\\
 &0, i \Gamma(1,0,0,1,0,1,1,0) + (\lambda- \kappa), \Gamma(0,0,1,1,1,0,1,0) )\\ 
 \underline{\Lambda}_8=&( 0,0,\Gamma(0,0,1,1,1,0,1,0),0,0,i\Gamma(0,0,0,0,1,1,1,1) - (\kappa + \lambda),0,\Gamma(a_2, a_2^\dagger)- \Gamma(a_2^\dagger, a_2), - \omega_2,0)\\
 \underline{\Lambda}_9=&(0,0, i \Gamma(0,1,1,0,1,0,0,1) + (\lambda-\kappa), \Gamma(0,0,1,1,1,0,1,0),0, \Gamma(1,0,1,0,0,0,1,1),\\
 &i\Gamma(0,0,0,0,1,1,1,1) - (\kappa + \lambda),\omega_2, 2 \Gamma(a_2, a_2^\dagger)-2 \Gamma(a_2^\dagger, a_2),-\omega_2   )\\
 \underline{\Lambda}_{10}=& (0,0,0,i \Gamma(0,1,1,0,1,0,0,1)+ (\lambda-\kappa),0,0,\Gamma(1,0,1,0,0,0,1,1),0, \omega_2,\Gamma(a_2, a_2^\dagger)- \Gamma(a_2^\dagger, a_2)  ).\\
\end{aligned}
\end{equation}
\end{widetext}
for both the matrix $\underline{\Lambda}$ and steady-state vector $\vec{G}$.
Solving this system of equations (numerically) leads to the symplectic covariance matrix. The actual covariance matrix is  obtained after symplectic transformation: $\sigma_{xixj}=\bar{\sigma}_{pipj}/2,~\sigma_{pipj}=\bar{\sigma}_{xixj}/2,~ \sigma_{xjpi}=-\bar{\sigma}_{xipj}/2$.
The steady-state occupation numbers are finally calculated  via $\langle a_1^\dagger a_1\rangle_{\text{ss}}=(\sigma_{x1x1}+\sigma_{p1p1} -1)/2$.

\section*{Appendix B: Quantum mutual information}
 The quantum mutual information for a Gaussian system can be calculated from the covariance matrix as $I(\sigma)= f(a)+f(b) -f(n_-(\sigma)) -f(n_+(\sigma))$  \cite{ser04}, with $a=\sqrt{\det(\alpha)},~b=\sqrt{\det \beta}$, $f(x)=(x+1/2)\ln(x+1/2)-(x-1/2)\ln(x-1/2)$, $n_{\mp}(\sigma)=\sqrt{\left(\Delta(\sigma)\mp\sqrt{\Delta(\sigma)^2-4\det  \sigma}\right)/2}$, $\Delta(\sigma)= \det\alpha+ \det \beta + 2 \det \gamma$, for the covariance matrix defined as $\sigma_{ij}=\langle x_i x_j+ x_j x_i \rangle/2$, $x_i=(x_1,p_1,x_2,p_2)$. In this form, the sub-matrices of interest are $\sigma=((\alpha, \gamma),(\gamma^T,\beta))$.

\section*{Appendix C: Quantum Langevin equations}
The quantum Langevin equation is  derived in the Heisenberg picture \cite{gar04}. This approach has the advantage that it does not involve  strong approximations as is the case for  quantum master equations \cite{pur16,riv10,boy17,lud10}. On the other hand, the drawback is that it cannot  easily   be solved in general as the corresponding differential equations are operator  differential equations in Hilbert space. For Gaussian systems,  it can however be solved  using matrix methods \cite{gar04}. The steady-state solution can thus be obtained by matrix inversion in Fourier space \cite{bre13}, $M(\nu) \vec{a}(\nu) + \vec{a}_\text{in}(\nu)=0$, with the two vectors $\vec{a}(\nu)= (\tilde{a}_1(\nu),\tilde{a}_1^\dagger (-\nu),\tilde{a}_2(\nu),\tilde{a}_2^\dagger (-\nu))$ and $\vec{a}_\text{in}(\nu)= (\sqrt{2\gamma_1}\tilde{a}_\text{1,in}(\nu)\sqrt{2\gamma_1}\tilde{a}_\text{1,in}^\dagger (-\nu),\sqrt{2\gamma_2}\tilde{a}_\text{2,in}(\nu),\sqrt{2\gamma_2}\tilde{a}_\text{2,in}^\dagger (-\nu))$. The matrix $M$ is explicitly given by  
\begin{equation}
M(\nu) = \left(\begin{matrix}
-i \nu + i \omega_1 &0 &i \kappa &i \lambda\\
0& -i\nu-i\omega_1 & -i\lambda & -i\kappa\\
i\kappa& i\lambda& -i\nu + i \omega_2  & 0\\
-i\lambda & -i\kappa& 0 & -i\nu-i\omega_2 
\end{matrix}\right)+ \bar{\gamma}
\end{equation}
with $\bar{\gamma}=\text{diag}(\gamma_1,\gamma_1,\gamma_2,\gamma_2)$. Inverting $M^{-1}=m$, the second moment $\langle a_1^\dagger a_1 \rangle$ in the algebraic space is
\begin{eqnarray}
\langle a_1^\dagger a_1 \rangle =&\int_{-\infty}^\infty \int_{-\infty}^\infty \langle a_1^\dagger(\nu) a_1(\nu')\rangle e^{i(\nu -\nu')t}/2\pi \text{d} \nu\text{d}\nu\nonumber\\
=&(1/\pi) \int_{-\infty}^\infty \gamma_1 (|m_{11}|^2 N(\nu,\beta_1) \nonumber\\
&+ |m_{12}|^2 (N(-\nu,\beta_1)+1))+\gamma_2 (|m_{13}|^2 N(\nu,\beta_2) \nonumber\\
&+ |m_{14}|^2 (N(-\nu,\beta_2)+1))\text{d}\nu
\label{Laneq}
\end{eqnarray}
and similar expressions for all the other second moments.

{The Gibbs state expectation values may  in addition be evaluated by the diagonalization of the Hamiltonian given above. In general, any quadratic expectation value in the $a_i$ algebraic space may be calculated via $\langle A_i A_j\rangle = \sum_{k\ell}S_{ik}S_{j\ell} \langle C_k C_\ell \rangle$ with $A_i = (a_1,a_2, a_1^\dagger, a_2^\dagger)$ and $C_i=(c_1,c_2,c_1^\dagger, c_2^\dagger)$. The $\langle C_i C_j\rangle$ are then given by the uncoupled  oscillators with eigenfrequencies, Eq.~\eqref{e2} of the main text, and corresponding temperature $T$.}

 \section*{Appendix D: Nonequilibrium steady states}
\begin{figure}[t]
\includegraphics[width=0.45\textwidth]{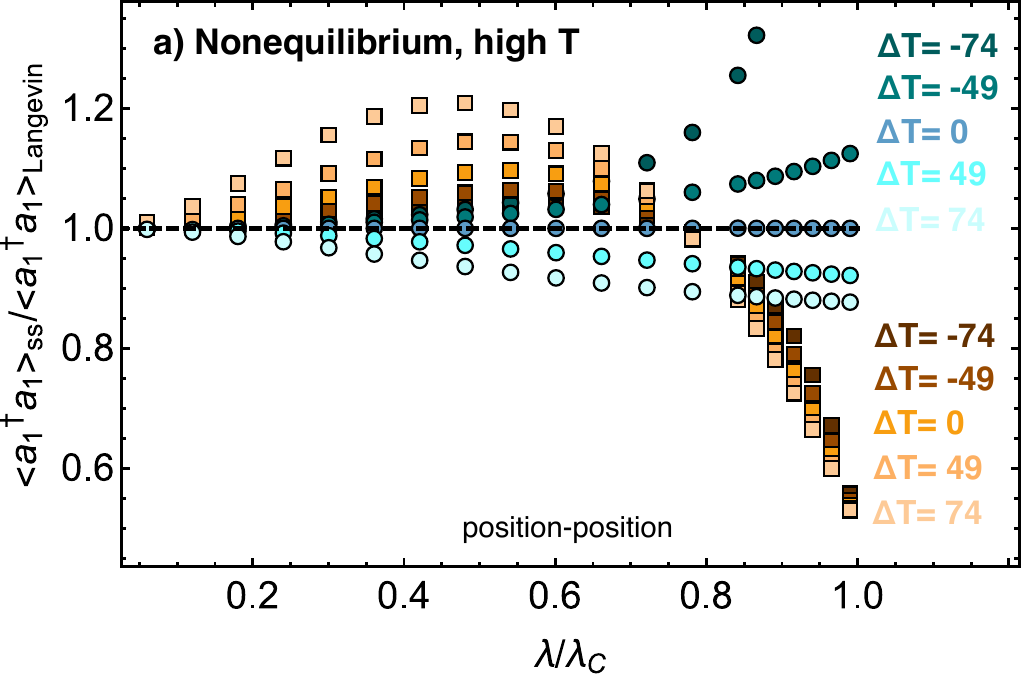}\\
\includegraphics[width=0.45\textwidth]{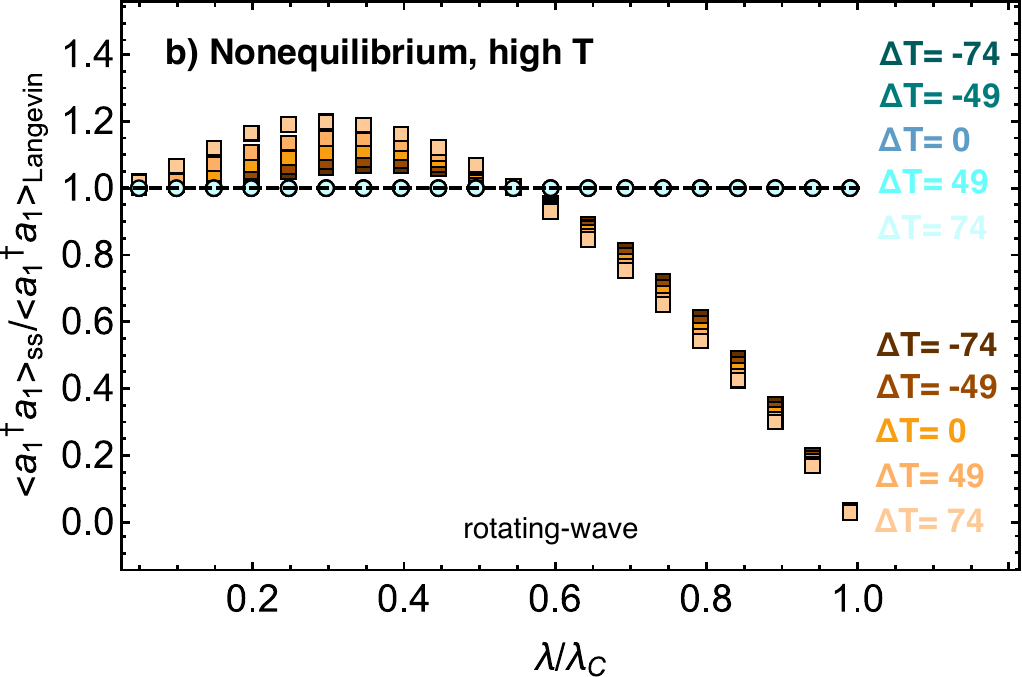}
\caption{Ratio  of the steady-state mean occupation numbers $\langle a_1^\dagger a_1\rangle_\text{ss}/ \langle a_1^\dagger a_1\rangle_\text{Langevin}$ of the quantum master equation and the quantum Langevin equation as a function of $\lambda/\lambda_\text{c}$, for various nonequilibrium temperature differences $\Delta T$, for a) position-position and b) rotating-wave interactions. Temperatures,  for positive  $\Delta T>0$, are $T_1=98, T_2= T_1 + \Delta T $, and for negative $\Delta T<0$, $T_1=T_2-\Delta T, T_2=  98$.}
 \label{fsgen}
 \end{figure}

 The deviation of the local (and, to a lesser extent, global) quantum master equations from the quantum Langevin equation does not only depend on the magnitude of the temperature difference $\Delta T$ but also on its sign (Figs.~\ref{fsgen}ab). We first note that 
  the local master equation  leads to   larger (smaller) mean occupation numbers for weak (strong) coupling, both for the position-position and the rotating-wave interactions. This  increase of the mean occupation number at small coupling is caused by the frequency difference between the  oscillators ($\omega_1>\omega_2$), which leads to a relatively larger occupation number in the second oscillator (modulated by the temperature difference), whereas its decrease  is induced by  strong-coupling effects.
In addition, the global master equation  completely matches the Langevin equation, for all $\Delta T$ for the rotating-wave interaction, while  this is not the case for the position-position interaction: the mean occupation number is larger (smaller) than that the quantum Langevin for $\Delta T <0$ ($\Delta T >0$).


\begin{thebibliography}{99}

    \bibitem{pau28}  W. Pauli, \"Uber das H-Theorem vom Anwachsen der Entropie vom Standpunkt der neuen Quantenmechanik, in Festschrift zum 60. Geburtstage A. Sommerfeld  (Hirzel, Leipzig 1928), p. 30.
\bibitem{bre02} H.-P. Breuer, and F. Petruccione, \textit{The Theory of Open Quantum Systems}, (Oxford University Press, Oxford, 2002).
\bibitem{gar04} C. Gardiner and P. Zoller, \textit{Quantum Noise}, (Springer, Berlin, 2004).
\bibitem{ali07}   R. Alicki and K. Lendi, \textit{Quantum Dynamical Semigroups and Applications}, (Springer, Berlin, 2007).
\bibitem{riv12} A. Rivas and S. F. Huelga, \textit{Open Quantum Systems}, (Springer, Berlin, 2012).
\bibitem{car93}   H. Carmichael, \textit{An Open Systems Approach to Quantum Optics} (Springer, Berlin, 1993).
 \bibitem{wei08} U. Weiss, \textit{Quantum Dissipative Systems}, (World Scientific, Singapore, 2008).  
 \bibitem{zwa01} R. Zwanzig, \textit{Nonequilibrium Statistical Mechanics}, (Oxford University Press, Oxford, 2001).
 \bibitem{nie00} M. A. Nielsen and  I. L. Chuang, \textit{Quantum Computation and Quantum Information}, (Cambridge
University Press, Cambridge, 2000).



\bibitem{die08} S. Diehl S, A. Micheli, A. Kantian, B. Kraus, H. P. B\"uchler and P. Zoller,
Quantum states and phases in driven open quantum systems with cold atoms, Nature Phys. \textbf{4}, 878 (2008).
\bibitem{die10} S. Diehl, A. Tomadin, A. Micheli, R. Fazio, and P. Zoller, Dynamical Phase Transitions and Instabilities in Open Atomic Many-Body Systems, Phys. Rev. Lett. \textbf{105}, 015702 (2010).
\bibitem{bre13} F. Brennecke, R. Mottl, K. Baumann, R. Landig, T. Donner, and T. Esslinger, Real-time observation of fluctuations at the driven-dissipative Dicke phase transition, Proc. Natl. Acad. Sci. USA \textbf{110}, 11763 (2013).
\bibitem{car13} C. Carr, R. Ritter, C. G. Wade, C. S. Adams, and K. J. Weatherill, Nonequilibrium Phase Transition in a Dilute Rydberg Ensemble, Phys. Rev. Lett. \textbf{111}, 113901 (2013).
\bibitem{mar14} M. Marcuzzi, E. Levi, S. Diehl, J. P. Garrahan, and I. Lesanovsky,
Universal Nonequilibrium Properties of Dissipative Rydberg Gases, Phys. Rev. Lett. \textbf{113}, 210401 (2014).
\bibitem{lab16} R. Labouvie, B. Santra, S. Heun, and H. Ott, Bistability in a Driven-Dissipative Superfluid, Phys. Rev. Lett. \textbf{116}, 235302 (2016).
\bibitem{sor18} M. Soriente, T. Donner, R. Chitra, and O. Zilberberg, Dissipation-Induced Anomalous Multicritical Phenomena, Phys. Rev. Lett. \textbf{120}, 183603 (2018).

\bibitem{pro08} T. Prosen and I. Pizorn, Quantum Phase Transition in a Far-from-Equilibrium Steady State of an 
XY Spin Chain, Phys. Rev. Lett. \textbf{101}, 105701 (2008).
\bibitem{kar09} D. Karevski and T. Platini, Quantum Nonequilibrium Steady States Induced by Repeated Interactions,
Phys. Rev. Lett. \textbf{102}, 207207 (2009).
\bibitem{car09} I. Carusotto, D. Gerace, H. E. Tureci, S. De Liberato, C. Ciuti, and A. Imamoglu, Fermionized Photons in an Array of Driven Dissipative Nonlinear Cavities, Phys. Rev. Lett. \textbf{103}, 033601 (2009).
\bibitem{pro10} T. Prosen and M. Znidaric, Long-Range Order in Nonequilibrium Interacting Quantum Spin Chains, Phys. Rev. Lett. \textbf{105}, 060603 (2010).
\bibitem{pro11} T. Prosen, Open XXZ Spin Chain: Nonequilibrium Steady State and a Strict Bound on Ballistic Transport, Phys. Rev. Lett. \textbf{106}, 217206 (2011).
\bibitem{pro11a} T. Prosen, Exact Nonequilibrium Steady State of a Strongly Driven Open XYZ Chain, Phys. Rev. Lett. \textbf{107}, 137201 (2011).
\bibitem{pro11b} T. Prosen and E. Ilievski, Nonequilibrium Phase Transition in a Periodically Driven XY Spin Chain,
Phys. Rev. Lett.  \textbf{107}, 060403 (2011).
\bibitem{vog12} M. Vogl, G. Schaller, and T. Brandes, Criticality in Transport through the Quantum Ising Chain, Phys. Rev. Lett. \textbf{109}, 240402 (2012).
\bibitem{cui15} J. Cui, J. I. Cirac, and M. C. Banuls,
Variational Matrix Product Operators for the Steady State of Dissipative Quantum Systems, Phys. Rev. Lett. \textbf{114}, 220601 (2015).
\bibitem{fos17} M. Foss-Feig, J. T. Young, V. V. Albert, A. V. Gorshkov, and M. F. Maghrebi, Solvable Family of Driven-Dissipative Many-Body Systems, Phys. Rev. Lett. \textbf{119}, 190402 (2017).
\bibitem{car19} F.  Carollo, E. Gillman, H. Weimer, and I. Lesanovsky, Critical Behavior of the Quantum Contact Process in One Dimension
Phys. Rev. Lett. \textbf{123}, 100604 (2019).
\bibitem{pop20}V. Popkov, T. Prosen, and L. Zadnik, Exact Nonequilibrium Steady State of Open XXZ/XYZ  Spin-1/2 Chain with Dirichlet Boundary Conditions,  Phys. Rev. Lett. \textbf{124}, 160403 (2020).
 
\bibitem{lev14} A. Levy and R. Kosloff, The local approach to quantum transport may violate the second law of thermodynamics, EPL \textbf{107}, 20004 (2014).

\bibitem{wal70} D. Walls, Higher order effects in the master equation for coupled systems, Z. Physik \textbf{234}, 231 (1970).
 \bibitem{car73} H. J.  Carmichael and D. F. Walls, Master equation for strongly interacting systems, J. Phys. A \textbf{6}, 1552 (1973).
 \bibitem{man15} P. D. Manrique, F. Rodriguez, L. Quiroga, and N. F. Johnson, Nonequilibrium Quantum Systems: Divergence between Global and Local Descriptions, Adv. Condens. Matter Phys. \textbf{2015},  615727 (2015).


 \bibitem{tru16} A. S. Trushechkin and I. V. Volovich, Perturbative treatment of inter-site couplings in the local description of open quantum networks, EPL \textbf{113}, 30005 (2016).
 \bibitem{gon17} J. O. Gonzalez, L. A. Correa, G. Nocerino, J. P. Palao, D.  Alonso and G. Adesso,  Testing the validity of
the 'local' and 'global' GKLS master equations on an exactly solvable model, Open Syst. Inf. Dyn. \textbf{24}, 1740010 (2017).
\bibitem{hof17} P. P. Hofer, M. Perarnau-Llobet, L. D. M. Miranda, G. Haack, R. Silva, J. B.  Brask, and N. Brunner, Markovian master equations for quantum thermal machines: Local versus global approach, New J. Phys. \textbf{19}, 123037 (2017).
\bibitem{sto17} J. T. Stockburger and T. Motz, Thermodynamic deficiencies of some simple Lindblad operators: a diagnosis and a suggestion for a cure, Fortschr. Phys. \textbf{65}, 1600067 (2017).
 \bibitem{nas18} M. T. Naseem, A. Xuereb, O E. Mustecaplioglu, Thermodynamic consistency of the optomechanical master equation, Phys. Rev. A \textbf{98}, 052123 (2018).
 \bibitem{chi17} G. De Chiara, G. Landi, A. Hewgill, B. Reid, A. Ferraro, A. J. Roncaglia and M. Antezza, Reconciliation of quantum local master equations with thermodynamics,  New J. Phys. \textbf{20}, 113024 (2018).
 \bibitem{mit18} M. T. Mitchison and M. B. Plenio,  Non-additive dissipation in open quantum networks out of equilibrium,  New J. Phys. \textbf{20}, 033005 (2018).
\bibitem{cat19} M. Cattaneo, G. L. Giorgi, S. Maniscalco, and R. Zambrini,
 Local versus global master equation with common and separate baths: superiority of the global approach in partial secular approximation, New J. Phys.  \textbf{21}, 113045 (2019).
 

 \bibitem{wic07} H. Wichterich, M. J. Henrich, H.-P. Breuer, J. Gemmer, and M. Michel,  Modeling heat transport through completely positive maps, Phys. Rev. E \textbf{76}, 031115 (2007).
 \bibitem{pur16} A. Purkayastha, A. Dhar, and M. Kulkarni, Out-of-equilibrium open quantum systems: A comparison of approximate quantum master equation approaches with exact results, Phys. Rev. A \textbf{93}, 062114 (2016).
\bibitem{riv10} A. Rivas, A. Douglas, K. Plato, S. F. Huelga and M. B Plenio, Markovian master equations: a critical study, New J. Phys. \textbf{12}, 113032 (2010).
 \bibitem{boy17} D. Boyanovsky and D. Jasnow, Heisenberg-Langevin versus quantum master equation, Phys. Rev. A \textbf{96}, 062108 (2017).
  \bibitem{lud10} M. Ludwig, K. Hammerer, and F. Marquardt, Entanglement of mechanical oscillators coupled to a nonequilibrium environment, Phys. Rev. A \textbf{82}, 012333 (2010).
 
 \bibitem{asp14} M. Aspelmeyer, T. J. Kippenberg, and F. Marquardt, Cavity optomechanics, Rev. Mod. Phys. \textbf{86}, 1391 (2014).
 \bibitem{dic54} R. H. Dicke, Coherence in Spontaneous Radiation Processes, Phys. Rev. \textbf{93}, 99 (1954).
 \bibitem{tav68} M. Tavis and F. W. Cummings, Exact Solution for an 
$N$-Molecule-Radiation-Field Hamiltonian, Phys. Rev. \textbf{170}, 379 (1968).
 \bibitem{bra05} T. Brandes, Coherent and collective quantum optical effects in mesoscopic systems, Phys. Rep. \textbf{408},  315 (2005).
 \bibitem{kir19} P. Kirton, M. M. Roses, J. Keeling, and E. G. Dalla Torre, Introduction to the Dicke Model: From Equilibrium
to Nonequilibrium, and Vice Versa, Advanced Quantum Technologies \textbf{2}, 1970013 (2019).
  \bibitem{ema03} C. Emary, and T. Brandes, Chaos and the quantum phase transition in the Dicke model,  Phys. Rev. E \textbf{67}, 066203 (2003).
\bibitem{lam04} N. Lambert, C. Emary, and T. Brandes, Entanglement and the phase transition in single mode superradiance, Phys. Rev. Lett. \textbf{92}, 7 (2004). 
\bibitem{sud12} V. Sudhir, M. G. Genoni, J. Lee, and M. S. Kim, Critical behavior in ultrastrong-coupled oscillators, Phys. Rev. A \textbf{86}, 012316 (2012).
\bibitem{hua20} J. F. Huang, J. Q. Liao, and L. M. Kuang, Ultrastrong jaynes-cummings model, Phys. Rev. A 101(4), 043835
(2020).
 \bibitem{for19} P. Forn-Diaz, L. Lamata, E. Rico, J. Kono, and E. Solano, Ultrastrong coupling regimes of light-matter interaction, Rev. Mod. Phys. \textbf{91}, 025005 (2019).
 
 \bibitem{hep73} K. Hepp and E. Lieb, On the superradiant phase transition for molecules in a quantized radiation field: the Dicke maser model, Ann. Phys. \textbf{76}, 360 (1973). 
\bibitem{hio73} Y. K. Wang and F. T. Hioe, Phase Transition in the Dicke Model of Superradiance, Phys. Rev. A \textbf{7}, 831 (1973).  
\bibitem{car73a} H. Carmichael, C. Gardiner, and D. Walls,  Higher order corrections to the Dicke superradiant phase transition, Phys. Lett. A \textbf{46},  47 (1973).
\bibitem{sup} See Appendix.
\bibitem{anf05}  A. Anfossi, P. Giorda, A. Montorsi, and F. Traversa, Two-point versus multipartite entanglement in quantum phase transitions, Phys. Rev. Lett. \textbf{95}, 056402 (2005).
\bibitem{sin11} R. R. P. Singh, M. B. Hastings, A. B. Kallin, and R. G. Melko, Finite-Temperature Critical Behavior of Mutual
Information, Phys. Rev. Lett. \textbf{106}, 135701 (2011).
\bibitem{alc13} F. C. Alcaraz and M. A. Rajabpour, Universal behavior of the Shannon mutual information of critical quantum chains, Phys. Rev. Lett. \textbf{111}, 017201 (2013).
\bibitem{tom17} G. De Tomasi, S. Bera, J. H. Bardarson, and F. Pollmann, Quantum Mutual Information as a Probe for Many-Body Localization, Phys. Rev. Lett. \textbf{118}, 016804 (2017).
\bibitem{wal19} C. Walsh, P. S\'emon, D. Poulin, G. Sordi, A. M. S.  Tremblay, Local entanglement entropy and mutual information across the Mott transition in the two-dimensional Hubbard model, Phys. Rev. Lett. \textbf{122},  067203 (2019).
\bibitem{Cam16} S. Campbell, G. De Chiara, M. Paternostro, Equilibration and nonclassicality of a double-well potential, Sc. Rep. \textbf{6}, 19730 (2016)
\bibitem{ser04} A. Serafini, F. Illuminati, S. De Siena, Symplectic invariants, entropic measures and correlations of Gaussian states, J. Phys. B: At. Mol. Opt. Phys \textbf{37} L21-L28 (2004)



\end{thebibliography}
\end{document}